\documentclass{article}

\usepackage{amsmath, amssymb}
\usepackage{fullpage}
\usepackage{listings}
\usepackage{url}
\usepackage{authblk}

\title{Simpler Online Updates for Arbitrary-Order Central Moments}
\author{Xiangrui Meng\thanks{meng@databricks.com}}
\affil{Databricks}
\date{}

\begin{document}

\maketitle

\begin{abstract}
  Statistical moments are widely used in descriptive statistics.
  Therefore efficient and numerically stable implementations are important in practice.
  P{\'e}bay \cite{pebay2008formulas} derives online update formulas for arbitrary-order central
  moments.
  We present a simpler version that is also easier to implement.
\end{abstract}

\noindent Summary statistics are commonly used in describing a data set.
For streaming or large-scale data, one-pass algorithms are preferred in practice.
Na{\"i}ve one-pass algorithms that update the moments directly are numerically unstable.
This can be easily seen from a data set with a small variance but a huge mean.
Stable update formulas via central moments were studied for variance \cite{welford1962note},
skewness, and kurtosis \cite{terriberry2007computing}.
P{\'e}bay \cite{pebay2008formulas} derived online update formulas for arbitrary-order central
moments.
In this note, we present a simpler version for updating central moments, which is also easier
to implement.

Given a sequence of values $\{x_i\}_{i=1,\ldots}$, let $M_{p,n} = \sum_{i=1}^n (x_i-\mu_n)^p$,
where $p \geq 2$ and $\mu_n = \sum_{i=1}^n x_i / n$ is the mean value.
Then the $p$-th central moment is $M_{p,n}/n$.
Define $\delta = x_n - \mu_{n-1}$, where $\mu_{n-1} = \sum_{i=1}^{n-1} x_i/(n-1)$.
It is easy to see that $\mu_n = \mu_{n-1} + \frac{\delta}{n}$, and we have
\begin{equation*}
\begin{aligned}
  M_{p,n} - M_{p,n-1}
  &= \sum_{i=1}^n (x_i - \mu_n)^p  - \sum_{i=1}^{n-1} (x_i - \mu_{n-1})^p \\
  &= \sum_{i=1}^n \left( (x_i - \mu_n)^p - (x_i - \mu_{n-1})^p \right) + (x_n - \mu_{n-1})^p \\
  &= \sum_{i=1}^n \left( (x_i - \mu_n)^p - \left( (x_i - \mu_n) + \frac{\delta}{n} \right)^p \right) + \delta^p \\
  &= - \sum_{i=1}^n \left( \sum_{k=1}^{p} \binom{p}{k} \left( \frac{\delta}{n} \right)^{k} (x_i - \mu_n)^{p-k} \right) + \delta^p \\
  &= - \sum_{k=1}^{p-1} \left( \binom{p}{k} \left( \frac{\delta}{n} \right)^{k} \sum_{i=1}^n (x_i - \mu_n)^{p-k} \right) - \sum_{i=1}^n \left( \frac{\delta}{n} \right)^p + \delta^p \\
  &=- \sum_{k=1}^{p-1} \binom{p}{k} \left( \frac{\delta}{n} \right)^{k} M_{p-k,n} + \delta \left(\delta^{p-1} - \left(\frac{\delta}{n}\right)^{p -1} \right),
\end{aligned}
\end{equation*}
Note that $M_{1,n} = 0$. So we get the following update rules:
\begin{equation}
  \begin{aligned}
  M_{2,n} &= M_{2,n-1} + \delta \left(\delta - \frac{\delta}{n}\right), \\
  M_{p,n} &= M_{p,n-1}  - \sum_{k=1}^{p-2} \binom{p}{k} \left( \frac{\delta}{n} \right)^{k} M_{p-k,n} + \delta \left(
    \delta^{p-1} - \left(\frac{\delta}{n}\right)^{p -1} \right), \quad \forall p > 2,\ n > 0.
  \end{aligned}
\end{equation}
The updates could be done in-place sequentially from lower-order moements to higher-order ones.
This is simpler than the one proposed in \cite{pebay2008formulas}, and the formula contains only
constants, $\delta$, $\delta/n$, and their powers in the coefficients, which is easier to implement.
When $p = 4$, we have
\begin{equation*}
\begin{aligned}
M_{2,n} &= M_{2,n-1} + \delta \left( \delta - \frac{\delta}{n} \right), \\
M_{3,n} &= M_{3,n-1} - 3 \frac{\delta}{n} M_{2,n} + \delta \left( \delta^2 - \left(\frac{\delta}{n}\right)^2 \right), \\
M_{4,n} &= M_{4,n-1} - 4 \frac{\delta}{n} M_{3,n} - 6 \left( \frac{\delta}{n} \right)^2 M_{2,n} +
\delta \left( \delta^3 - \left(\frac{\delta}{n}\right)^3 \right).
\end{aligned}
\end{equation*}
By storing $\delta / n$, only one division is needed per update. For example, the following
implementation in Python uses 26 floating-point operations (FLOPs) per update, including one
division.
\begin{lstlisting}
n += 1
delta = x - mu
delta_n = delta / n
mean += delta_n
m2 += delta * (delta - delta_n)
delta_2 = delta * delta
delta_n_2 = delta_n * delta_n
m3 += -3.0 * delta_n * m2 + delta * (delta_2 - delta_n_2)
m4 += -4.0 * delta_n * m3 - 6.0 * delta_n_2 * m2 \
  + delta * (delta * delta_2 - delta_n * delta_n_2)
\end{lstlisting}
This is slightly better (less FLOPs) than the one proposed in \cite{terriberry2007computing},
and it is easier to extend to higher-order moments.
As a final note, the merge formula for parallel updates can be found in
\cite{terriberry2007computing} and \cite{pebay2008formulas}, and the standardized moments could be
easily derived from $M_{p,n}$, for example,
\begin{eqnarray*}
\text{variance} &=& M_{2,n} / n, \\
\text{skewness} &=& n^{1/2} M_{3,n} / M_{2,n}^{3/2}, \\
\text{kurtosis} &=& n M_{4,n} / M_{2,n}^2.
\end{eqnarray*}

\bibliographystyle{plain}
\bibliography{simpler-moments}

\end{document}